\documentstyle[11pt]{article}
\begin{document}
\thispagestyle{empty}
\baselineskip 18pt
\rightline{SNUTP-99-012}
\rightline{KIAS-P99015}
\rightline{{\tt hep-th}/9903095}

\

\def\tr{{\rm tr}\,} \newcommand{\beq}{\begin{equation}}
\newcommand{\eeq}{\end{equation}} \newcommand{\beqn}{\begin{eqnarray}}
\newcommand{\eeqn}{\end{eqnarray}} \newcommand{\bde}{{\bf e}}
\newcommand{\balpha}{{\mbox{\boldmath $\alpha$}}}
\newcommand{\bsalpha}{{\mbox{\boldmath $\scriptstyle\alpha$}}}
\newcommand{\bbeta}{{\mbox{\boldmath $\beta$}}}
\newcommand{\bgamma}{{\mbox{\boldmath $\gamma$}}}
\newcommand{\bsbeta}{{\mbox{\boldmath $\scriptstyle\beta$}}}
\newcommand{\blambda}{{\mbox{\boldmath $\lambda$}}}
\newcommand{\bsigma}{{\mbox{\boldmath $\sigma$}}}
\newcommand{\bslambda}{{\mbox{\boldmath $\scriptstyle\lambda$}}}
\newcommand{\ggg}{{\boldmath \gamma}} \newcommand{\ddd}{{\boldmath
\delta}} \newcommand{\mmm}{{\boldmath \mu}}
\newcommand{\nnn}{{\boldmath \nu}}

\newcommand{\bra}[1]{\langle {#1}|}
\newcommand{\ket}[1]{|{#1}\rangle}
\newcommand{\sn}{{\rm sn}}
\newcommand{\cn}{{\rm cn}}
\newcommand{\dn}{{\rm dn}}
\newcommand{\diag}{{\rm diag}}

\

\vskip 1cm
\centerline{\Large\bf Massless Monopoles and Multi-Pronged Strings}
\vskip 0.2cm

\vskip 1.2cm
\centerline{\large\it
Kimyeong Lee\footnote{Electronic Mail: kimyeong@phya.snu.ac.kr}}
\vskip 3mm 

\centerline{\large \it Physics Department and Center for Theoretical Physics}
\centerline{\large \it Seoul National University, Seoul 151-742, Korea}

\vskip 1.5cm

\vskip 2mm
\begin{quote}
{\baselineskip 16pt We investigate the role of massless magnetic
monopoles in the $N=4$ supersymmetric Yang-Mills Higgs theories.  They
can appear naturally in the 1/4-BPS dyonic configurations associated
with multi-pronged string configurations.  Massless magnetic monopoles
can carry nonabelian electric charge when their associated gauge
symmetry is unbroken.  Surprisingly, massless monopoles can also
appear even when the gauge symmetry is broken to abelian subgroups.}
\end{quote}


\newpage

Recently there has been several activities in trying to understand
multi-pronged strings ending on parallel D3 branes in the type IIB string
theory~\cite{bergman}. 1/2 BPS monopoles and W-bosons in $N=4$
supersymmetric Yang-Mills theories appear naturally as D-strings and
fundamental strings connecting parallel D3 branes in type IIB string
theory~\cite{witten}.  1/4-BPS multi-pronged string webs have a
natural interpretation in the N=4 supersymmetric Yang-Mills theories:
they correspond to the configurations of many dyons separated in some
distance, balanced under the electromagnetic Coulomb force and more
complicated Higgs force~\cite{leeyi}. When there are several dyons,
the configuration can be spherically symmetric for some cases and more
explicit field configuration can be obtained~\cite{hasimoto}. The
number of zero modes of these configurations  is somewhat
complicated~\cite{bak,bergman}, than the 1/2 BPS monopole case~\cite{ejw}

For 1/4-BPS configurations of dyons, only two independent components
of six Higgs fields in $N=4$ supersymmetric Yang-Mills theory get
involved~\cite{leeyi}. In string picture, the eigenvalues of the
vacuum values of six Higgs fields indicate the positions of D2 branes
on the transverse 6-dimensional space. The 1/4-BPS configurations lies
on a plane, forming a string web~\cite{bergman}.  One component of the
Higgs fields determines the masses of magnetic monopoles, and some of
them can be massless.  As two independent Higgs asymptotic values
determine the symmetry breaking pattern, the gauge symmetry
corresponding to the massless monopole can be unbroken or broken.
When the gauge symmetry for the massless monopoles is not broken, they
can not exist in isolation. But in presence of other massive magnetic
monopoles, their presence can be felt. Especially they can carry
nonabelian electric charge.  When the gauge symmetry is completely
broken, the corresponding massless monopoles will become massive in
isolation. Thus, their appearance seems mysterious.

In this paper, we investigate in detail the physics of such massless
monopoles in 1/4-BPS dyonic configurations by employing a simple
model. The simple model we study is based on the configurations for
three distinct monopoles when the gauge symmetry is broken from
$SU(4)$ to $U(1)\times SU(2) \times U(1)$ or $U(1)^3$.  We consider
this model because it is easy to construct the BPS configurations by
the ADHMN method~\cite{adhmn} and the partially broken limit of the
configuration can be easily explored.

There are several fascinating problems associated with monopoles when
the unbroken gauge symmetry has a nonabelian subgroup.  When massive
magnetic monopoles carry nonabelian magnetic charge, it is well known
that one cannot define the electric nonabelian transformations
globally around them, which is called `the global color
problem'~\cite{global}. The main reason behind it is that the inertia
for some of global gauge transformations become infinite. This is
associated with the fact that the corresponding zero modes are
nonnormalizable.  When there are several magnetic monopoles so that
the net magnetic charge is purely abelian, then one can define the
global color transformations without such obstruction.

There is  also the issue of massless monopoles.  When the gauge
symmetry is partially restored from abelian group to nonabelian group,
some of magnetic monopoles become massless.  The idea of such
monopoles in isolation is meaningless: their size becomes infinite
and their field strength vanishes. However, it is known now that these
massless monopoles can leave some footprints when one start with
several other monopoles~\cite{lwy3,so5,dancer,lu1,wy}. The physics of
massless monopoles becomes especially simpler when the net magnetic
charge remain abelian.  In this case massless monopoles form magnetic
cloud surrounding massive magnetic monopoles, and so the net magnetic
charge is abelian. In the massless limit, the dimension of moduli
space does not change, but the characteristics of moduli parameters
for massless monopoles changes drastically. Some of them describe the
orbit of unbroken electric gauge transformations, and the rest do the
cloud shape parameters~\cite{lwy3}. When there are several identical
massless monopoles, some of them can be taken to the infinity, making
the net magnetic charge to be nonabelian. In this case the global
charge is not well defined, but there will be still some remaining
cloud parameters.

When we consider a single Higgs field case, the magnetic monopole
configuration describes  1/2-BPS states.  The moduli space dynamics of
the zero modes of three distinct fundamental monopoles,
when $SU(4)$ is broken to $U(1)^3$, has been studied in dynamics detail
in the moduli space approximation~\cite{lwy2}. Also, the moduli space
of these monopoles when one for the middle root of the Dynkin diagram
becomes massless has been studied~\cite{lwy3}.  More recently, the
detail field configuration of such a system~\cite{wy} have been
obtained by using the ADHMN construction~\cite{adhmn}. Its salient
feature is that four zero mode parameters  for the massless
monopole are composed of three $SU(2)$ gauge orbit parameters and one
gauge invariant parameter.  The nonabelian cloud surrounding the
massive monopoles takes an ellipsoid shape, at whose two focal points
lie two massive monopoles. The size of the ellipsoid is the gauge
invariant parameter.

As shown in Ref.~\cite{leeyi}, 1/4-BPS configurations for dyons have 
only two relevant components of six Higgs fields, $\phi_I$,
$I=1,...,6$.  The BPS energy bound is saturated when the several BPS
equations are satisfied. They can be put into two BPS equations,
\beqn
&& B_i = D_i b\cdot \phi,\\
&& D_i^2 a\cdot \phi - [b\cdot\phi,[b\cdot\phi,a\cdot \phi]]=0 ,
\eeqn
where $a_I$ and $b_I$ are two orthonormal vectors in 6 dimensions and
decided by the electric and magnetic charge vectors,
\beqn
&& Q^M_I = 2\int d^3x \partial_i ({\rm tr} B_i \phi_I) ,\\ 
&& Q^E_I = 2\int d^3x \partial_i ({\rm tr} E_i \phi_I) .
\eeqn
It was shown that the the second BPS equation is identical to the unbroken
gauge zero modes for the first BPS equation. The first equation is
solved in Ref.~\cite{wy}. To solve the second equation, we follow
the method in Ref.~\cite{leeyi}.

To start, we assume that the asymptotic value of $b\cdot \phi$ is
diagonalized to be
\beq
b\cdot \phi(\infty) = {\rm diag}(h_1,h_2,h_3,h_4) ={\bf h}\cdot {\bf  H},
\eeq
where we choose the gauge so that $h_1<h_2< h_3 <h_4$. The diagonal
values are the coordinate of D3 branes along the $b_I$ direction. In
this case the gauge symmetry is spontaneously broken to $U(1)^3$.
There exists a unique set of unit length simple roots, $\balpha,
\bbeta, \bgamma$ such that their inner product with ${\bf h} $ in the
root space is positive definite.  We can choose the order so that
\beqn
&& \balpha\cdot {\bf H} = {\rm diag}(-1/2,1/2,0,0), \\
&& \bbeta\cdot{\bf H} = {\rm diag} (0,-1/2,1/2,0), \\
&& \bgamma\cdot{\bf H} = {\rm diag}(0,0,-1/2,1/2).
\eeqn
The masses of $\balpha, \bbeta, $ and $\bgamma$ monopoles are $4\pi$ times
\beqn
&& \mu_1 = h_2-h_1 ,\nonumber \\
&& \mu_2 = h_3-h_2, \nonumber \\
&& \mu_3 = h_4-h_3,  
\eeqn
respectively. (Here we put the gauge coupling constant to be unity.)

We consider three distinct fundamental monopoles, whose magnetic charge is
\beq
{\bf g}= 4\pi (\balpha+\bbeta+\bgamma).
\eeq
In the limit $\mu_2=0$, the electric gauge symmetry is partially
restored to $U(1) \times SU(2)\times U(1)$.  In this limit the
$\bbeta$ monopole becomes massless and the above magnetic charge
remain abelian as ${\bf g}\cdot\bbeta = 0$.  From the first BPS
equation, the magnetic charge is related to the asymptotic form of
$b\cdot\phi$,
\beq
 b\cdot \phi = b\cdot\phi(\infty) -\frac{{\bf g}\cdot
 {\bf H}}{4\pi r} + {\cal O}\left(\frac{1}{r^2}\right).
\label{bphia}
\eeq
Similarly, the electric charge
\beq
{\bf q} = q_\alpha \balpha + q_\beta \bbeta +q_\gamma \bgamma
\eeq
is related to the  the asymptotic form of $a\cdot\phi$,
\beq
 a\cdot\phi = a\cdot\phi(\infty) - \frac{{\bf q}\cdot
 {\bf H}}{4\pi r} +  {\cal O}\left(\frac{1}{r^2}\right).
\eeq
One goal of this paper is to find the electric charge ${\bf q}$ for
the given magnetic charge configuration and $a\cdot\phi(\infty)$.
The asymptotic values $b\cdot\phi(\infty)$ and $a\cdot\phi(\infty)$
are the coordinate values of D3 branes along the orthogonal directions
$b_I$ an $a_I$.

The field configurations satisfying the first BPS equation has been
obtained by Weinberg and Yi in Ref.~\cite{wy}.  Let us review briefly
this derivation for the later sake.  For this purpose, we put the
position of three monopoles at
\beqn
&& {\bf r}_\alpha = (0,0, - \frac{\mu_3}{M} R), \\
&& {\bf r}_\beta =  (r\sin\theta, 0,  r\cos\theta), \\
&&  {\bf r}_\gamma = (0,0, \frac{\mu_1}{M} R),
\eeqn
where $M=\mu_1+\mu_3$. The origin coincides with the position of the
center of mass.  The relative positions of the  $\bbeta$
monopole with respect to two massive monopoles are
\beqn
{\bf y}_L = {\bf r}_\beta  -{\bf r}_\alpha, \nonumber \\
{\bf y}_R = {\bf r}_\beta - {\bf r}_\gamma .
\eeqn
We call their magnitude $y_L, y_R$. 
In the limit where $\bbeta$ monopole becomes massless, the unbroken
$SU(2)$ gauge invariant parameters are
\beqn
&& R= |{\bf r}_\alpha - {\bf r}_\gamma| , \\
&& D = y_L + y_R .
\eeqn
$R$ is the distance between two massive monopoles and $D$ is the cloud
parameter. For a given $D$, the position of the massless monopole becomes
two dimensional ellipsoidal surface.  As shown in Ref.~\cite{wy},
the nonabelian component of magnetic charge becomes dipole-like
outside this ellipsoid. Thus our configuration has a nonabelian
ellipsoidal cloud surrounding the Coulomb like nonabelian source such
that the total configuration is completely abelian. One could regard
that the nonabelian cloud as a reminiscent of the confining string of
 quarks in a meson of QCD.  Here it does not give a linear
potential, but neutralizes in the color magnetic charge.

From the asymptotic value of $b\cdot \phi$, we see three intervals
$[h_1,h_2]$, $[h_2,h_3]$ and $[h_3,h_4]$ for the parameter $t$ of the
Nahm equation.  The Nahm data, which is the solution of the Nahm equation, is
\beq
 {\bf T}(t) = \left(\begin{array}{ccc}
                -{\bf r}_\alpha & {\rm for} &  t\in  (h_1,h_2)  ,\\ 
              -{\bf r}_\beta  & {\rm for} &  t\in   (h_2,h_3), \\
              -{\bf r}_\gamma & {\rm for} &  t\in  (h_3,h_4). \\
              \end{array} \right.
\eeq
In the  Nahm data, there are also  two  complex row vectors
\beqn 
&& {\bf a}_1 = \sqrt{2y_L} (\cos\frac{\theta_L}{2}, \sin \frac{\theta_L}{2}) 
 \;\;\;\; {\rm at}\;\;\; t=h_2,
\\
&& {\bf a}_2 = \sqrt{2y_R}(\cos \frac{\theta_R}{2}, \sin \frac{\theta_R}{2}) 
\;\;\;\; {\rm at}\;\;\; t=h_3,
\eeqn
such that 
\beqn
&& \frac{1}{2} {\rm tr}\, \sigma^i {\bf a}_1^\dagger  {\bf a}_1 = {\bf
y}_L= y_L(\sin\theta_L, 0 , \cos\theta_L) ,\\
&& \frac{1}{2} {\rm tr} \, \sigma^i {\bf a}_2^\dagger {\bf a}_2 =
-{\bf y}_R  =y_R(\sin\theta_R,0,\cos\theta_R),
\eeqn
with $0\le \theta_L, \theta_R  < 2\pi$.

The ADHMN equation~\cite{adhmn} for constructing $b\cdot\phi$ is
\beq 
\left[ -\frac{d}{dt} + \bsigma \cdot ({\bf T}(t)+ {\bf r}) \right] v(t)
 + {\bf a}_2^\dagger S_2
\delta(t-h_2) + {\bf a}_3^\dagger  S_3 \delta(t-h_3) = 0, 
\label{adhmn1}
\eeq
where $v(t)$ is a $2\times 4$ matrix and $S_2, S_3$ are 4-dimensional
row  vectors. Sometimes, we regard $v(t)$ and $S_2,S_3$ as four
independent solutions of the above equation.  The normalization
condition is
\beq
I_{4\times 4} =  \int_{h_1}^{h_4} dt \,\, v^\dagger(t) v(t)
 + S^\dagger_2 S_2 + S^\dagger_3 S_3 .
\eeq

There are four independent solutions of $v(t)$ and $S_2,S_3$ of
Eq.~(\ref{adhmn1}), which form four columns.  The first two
independent solutions are continuous solutions with $S_2=S_3=0$. In
the limit $\mu_2=0$, $v(t)$ for two solutions can be put in a $2\times
2$ matrix,
\beq
v(t) = \left( \begin{array}{lrl}
                f_1(t) N^{-1/2} & {\rm for} & t\in (h_1,h_2), \\
                f_3(t) N^{-1/2} & {\rm for} & t\in (h_3,h_4), \\
		\end{array}
\right.
\eeq
where
\beqn
&& f_1(t) =e^{(t-h_2)({\bf r}-{\bf r}_\alpha) \cdot \bsigma },\nonumber \\
&& f_3(t)=  e^{(t-h_3)({\bf r}-{\bf r}_\beta)\cdot\bsigma },
\eeqn
and the normalization factor $N$ is  the sum 
\beq
N=N_L + N_R,
\eeq
with
\beqn
&& N_L = \int_{h_1}^{h_2}\;dt \, f_1(t)^2, \nonumber \\
&& N_R = \int_{h_3}^{h_4} \;dt \, f_3(t)^2 .
\eeqn

There are two additional independent solutions, which can be written together
as a $2\times 2$ matrix, 
\beq
v(t) = \left( \begin{array}{ll}
		f_1(t)\eta_1, & t\in (h_1,h_2) \\
		f_3(t) \eta_2, &t\in (h_3,h_4) \\
		\end{array}
\right. ,
\eeq
where
\beqn
&& \eta_1 = - N_L^{-1} M [ {\bf a}_2^\dagger S_2 + {\bf a}_3^\dagger S_3], \\
&& \eta_3= N_R^{-1} M [{\bf a}_2^\dagger S_2 + {\bf a}_3^\dagger S_3] ,
\eeqn
with
\beq
M = N_L^{-1} +N_R^{-1}.
\eeq
Here we regard the nonvanishing $S_2, S_3$ as two dimensional row  vectors.
They are fixed by normalization condition,
\beq
\delta^{\mu\nu} = S_i^\mu [ \delta_{ij} + {\bf a}_i M {\bf a}_j^\dagger] 
S_j^\nu,
\eeq
where  $i,j=2,3$ and $\mu, \nu=1,2$ are column indices. 

From these solutions for $v(t)$ and $S_2$ and $S_3$, one can find the
Higgs field of the first BPS equation by using the formula,
\beq
b\cdot\phi({\bf r})  = \int_{h_1}^{h_4} dt \, t \, v^\dagger(t)v(t) + h_2  
S_2^\dagger S_2 + h_3 S_3^\dagger S_3 .
\eeq
The characteristic of this field is discussed  in Ref.~\cite{wy}.
We note here that this solution has the desired asymptotic form
(\ref{bphia}), once a further gauge transformation in $SU(4)$ is made.

The second BPS equation can be solved by finding the zero modes
corresponding to the global part of the unbroken gauge group
$U(1)\times SU(2)\times U(1)$. We use the strategy used in
Ref.~\cite{leeyi}. In our context we need to find an ordinary function
$p(t)$ for a given traceless diagonal matrix ${\bf q}\cdot {\bf H} =
{\rm diag} (q_1,q_2,q_3,q_4)$.  $p(t)$ satisfies the ordinary equation
\beq
\ddot{p}(t) -  W(t) p(t) + \Lambda(t) = 0 ,
\eeq
where
\beqn
&& W(t) = \tr_2 {\bf a}^\dagger_1 {\bf a}_1 \delta (t-h_2) + \tr_2 
{\bf a}^\dagger_2  {\bf a}_2 \delta(t-h_3),\\ 
&& \Lambda(t) = q_2  \tr_2 {\bf a}^\dagger_1 {\bf a}_1 \delta (t-h_2) + 
q_3 \tr_2  {\bf a}^\dagger_2  {\bf a}_2 \delta(t-h_3).
\eeqn
The boundary conditions for $p(t)$ are 
\beq
p(h_1)= q_1 \;\;\; {\rm and}\;\;\;\; p(h_4) = q_4.
\eeq
This boundary conditions can be obtained easily by taking the limit of the 
additional monopole need for a single caloron to infinity~\cite{leeyi}.

In the limit $\mu_2=0$, the solution becomes considerably simplified.
The solution is
\beq
p(t) = \left(    \begin{array}{ll}
                  q_1 +b_1(t-h_1) & t\in (h_1,h_2)\\
                  q_1+b_1\mu_1 + b_3(t-h_3) & t\in (h_3,h_4)
\\
\end{array} \right. ,
\eeq
where
\beqn
&& b_1 = \frac{1}{\Delta}\left\{ - ( 1+2\mu_3 D) q_1 + 2\mu_3 
(y_L q_2 +y_R q_3)  + q_4  \right\}, \nonumber \\
&& b_3 = \frac{1}{\Delta}\left\{ 
-q_1 - 2\mu_1 ( y_L q_2 +  y_R q_3) + (1+2\mu_1 D) q_4 \right\},
\eeqn
with 
\beq
\Delta = \mu_1+\mu_3+2\mu_1\mu_3 D .
\eeq
Then
the solution for the second BPS equation is
\beq
\Lambda({\bf r}) = \int_{h_1}^{h_4} \; dt \; v(t)^\dagger p(t) v(t) + 
q_2 S_2^\dagger S_2 + q_3 S_3^\dagger S_3.
\label{Lambda}
\eeq

In this paper we are not interested in the detail structure of the
second Higgs field in finite region, which is associated with the fact
how D3 branes are deformed to have multi-pronged string shape.
Instead we are interested in only the asymptotic behavior of the
field $\Lambda({\bf r})$.  By taking the similar approach as in
Ref.~\cite{wy}, we can get the asymptotic form 
\beqn
\Lambda({\bf r}) = && {\rm diag} (q_1,q_2,q_3,q_4)  \nonumber \\
&&  \;\; + \frac{1}{2r} \left( 
\begin{array}{cccc}
		b_1 & 0 & 0 &  0 \\
		0 &  y_L (c-q_2+q_3) & \sqrt{y_L y_R} \, c  \cos 
\frac{\theta_L-\theta_R}{2} & 0 \\
0 & \sqrt{y_L y_R} \, c  \cos \frac{\theta_L-\theta_R}{2} &
 y_R (c+q_2 -q_3) & 0 \\
0 & 0 & 0 & -b_3 \\
\end{array} \right)  + {\cal O}\left(\frac{1}{r^2}\right),
\eeqn
where $c=2q_1+2\mu_1b_1 -q_2-q_3$. (The asymptotic behavior at spatial
infinity can be calculated along a particular direction, say, along a
region where $e^{\mu_1 y_L}/y_L >> e^{\mu_3 y_R}/y_R$. In this region,
terms with vanishing  entry is of order $1/r^2$ or exponentially small.) 
 One can easily show that this
hermitian matrix is traceless. This is the asymptotic form of the
general solution for the second BPS equation.

There are three independent choices of the charge ${\rm diag}
(q_1,q_2,q_3,q_4)$ such that $\sum_i q_i=0$, which lead to three
independent solutions, $\Lambda_1({\bf r})$, $\Lambda_2({\bf r})$,
$\Lambda_3({\bf r})$ for the second BPS equation.  $a\cdot\phi$ is in
general a linear combination of these three solutions. For simplicity,
we consider three cases where $a\cdot\phi$ to be proportional to
$\Lambda_i$ for each $i=1,2,3$.

First one is the obvious one,
\beq
\Lambda_1 =  b\cdot \phi({\bf r}) .
\eeq
In this case the asymptotic value $\Lambda_1(\infty) \sim b\cdot
\phi(\infty)$, so  that four $D3$ branes will lie on a line. The
electric charge will be proportional to the total magnetic charge. The
configuration becomes 1/2-BPS rather than 1/4-BPS.

The second case appears with the gauge symmetry being still broken to
$U(1)\times SU(2) \times U(1)$. However, parallel D3 branes do not lie
on a line.  In this case, the asymptotic form is
\beq
\Lambda_2 = {\rm diag}(  2\mu_3,  -\mu_1 -\mu_3,  -\mu_1-\mu_3,  2 \mu_1) 
-\frac{1}{4\pi r}( q_\alpha \balpha +q_\beta \bbeta + q_\gamma \bgamma)
\cdot {\bf H} +{\cal O}\left(\frac{1}{r^2}\right),
\eeq                            
where
\beqn
&& q_\alpha  = \frac{8\pi}{\Delta}\{ \mu_1-\mu_3-(\mu_1+3\mu_3)\mu_3 D\}, \\
&& q_\beta = \frac{4\pi}{\Delta}\{ 2(\mu_1-\mu_3) +3(\mu_1^2-\mu_3^2)D 
   \pm ( 3 \mu_1^2 +2\mu_1 \mu_3 + 3\mu_3^2)R\}, \\
&& q_\gamma  = \frac{8\pi}{\Delta}\{\mu_1-\mu_3 + (3\mu_1+\mu_3)\mu_1 D \} .
\eeqn
The electric charges depend only on the $SU(2)$ gauge invariant
parameters $D$ and $R$. There is a global part of the unbroken local
$SU(2)$ gauge group, which exchanges two middle diagonal components.
Under this, the sign of the  $R$ term in $q_\beta$ changes. (Notice also
that both $\balpha$ and $\bgamma$ electric charges are not invariant
under the $SU(2)_\beta$ gauge transformations.)

 When the cloud size is smallest so that $D=R$, $q_\beta$ is equal to
$q_\alpha$ or $q_\gamma$, depending on this sign. This is consistent with
the view in Ref.~\cite{wy} that the massless $\bbeta$ monopole lies on
the top of the $\balpha$ or $\bgamma$ monopole in the minimum cloud
size case.  In the large $D$ limit, there exist the critical amounts
of electric charges. Similar to the $SU(3)$ case, when the electric
charges exceed the critical amounts, there is no 1/4 BPS
configurations because the electric repulsion overcomes the Higgs
attraction. 

The electric charge does not need to vanish even when all three
monopoles are overlapped so that $D=R=0$ unless their core sizes are
identical $\mu_1=\mu_3$. When two massive monopoles are overlapped so
that $R=0$, note that ${\bf q}$ is purely abelian as ${\bf q}\cdot
\bbeta=0$.

The third case will be
\beq
\Lambda_3 = {\rm diag} (0,  -\mu_1-\mu_3,  \mu_1+\mu_3, 0)
-\frac{1}{4\pi r}\{q_\alpha \balpha +q_\beta \bbeta +q_\gamma \bgamma
\}\cdot {\bf H} +{\cal O}\left(\frac{1}{r^2}\right)
\eeq
where
\beqn
&& q_\alpha = -\frac{8\pi}{\Delta} \mu_3(\mu_1+\mu_3)(y_L-y_R)  \\
&& q_\beta =  \frac{8\pi}{\Delta}(\mu_1+\mu_3)\{\mu_1 y_L+\mu_3 y_R + 
4\mu_1\mu_3 y_L y_R \} \\
&& q_\gamma = \frac{8\pi}{\Delta}\mu_1(\mu_1+\mu_3)(y_L-y_R)
\eeqn
As $a\cdot\phi \sim \Lambda_3$, the gauge symmetry is completely
broken to $U(1)^3$.  Hence, there are  no massless monopoles in
isolation. Naturally, there are two questions arising here. First is
how to interpret the above result. Second is how to understand the
existence of massless monopoles in the configuration.

We note that the electric charges given in the above equations depend
on the position of the $\beta$ monopole. While the $\beta$ monopole
appears massless in the example, it position cannot be transformed by
unbroken abelian subgroup, contrast to the second case. However, the
electric charges shows the obvious symmetry along the rotation around
the line connecting two massive monopoles, as it depends only on $y_L,
y_R$. 

The reason for the $\beta$ monopole being massless seems obscure in
the field theoretic picture, while it is much more clear in the string
picture. Here, as one moves the positions of $D3$ branes
or changes electric charge, two vectors $b_I$ and $a_I$ change. The
genuine symmetry restoration occurs when two D3 branes overlap. On the
other hand,  a given magnetic monopole appears massless when the $b_I
\phi_I$ components for two D3 branes have identical value. 
Thus the line connecting corresponding two D3 branes is perpendicular
to the $b_I$ vector.

 To find why the $\beta$ monopole can appear massless in the field
theory picture, let us start from a configuration for three massive
$\alpha$, $\beta$ and $\gamma$ monopoles.  When there are two Higgs
expectation values which are not parallel, these monopoles are
attracted to each other. Let us put some electric charges such that
relative charges are not zero, and then three dyons have repulsive
force between them.  They will be settled, separated in some
distance. Also the core of dyons gets larger. When electric charges
are increasing, two vectors $a_I$ and $b_I$ are transforming. If the
condition is right, it seems that any one of them appears massless as
far as the first BPS equation is concerned.  Our case corresponds to
the case where the $\beta$ monopole becomes massless.

However, we will argue that the $\alpha$ or $\gamma$ monopoles cannot
appear massless in a finite region in our case.  For simplicity,
consider just two $\alpha$ and $\beta$ monopoles in the $SU(3)$ gauge
theory. If the $\alpha$ monopole becomes massless, the configuration
has nonzero nonabelian charge as far as the first BPS equation is
concerned. This configuration, however, has no nonabelian cloud, and
so the configuration for the first BPS equation is independent of the
position of the $\alpha$ monopole~\cite{lu2}. It seems that the
$\alpha$ monopole has disappeared. The solution to this paradox is
that as far as we stay in BPS configurations, the $\alpha$ monopole
can appear massless only when it goes to infinity. This can be easily
seen from the string picture. This is the Bergman's
instability~\cite{bergman} where the three pronged configuration
becomes unstable. In the field theoretic term, the $\alpha$ electric
charge (say without $\beta$ electric charge) keeps $\alpha$ and
$\beta$ monopoles separated in finite distance. Once we change the
asymptotic Higgs values so that the $\alpha$ monopole appears
massless, the electric repulsion sends the $\alpha$ monopole to
infinity. However, we can always have a massive $\alpha$ monopole in
finite distance once we consider non-BPS configurations.

Instead if we started with one $\alpha$ and two $\beta$ monopoles, the
$\alpha$ monopole can appear massless at a finite region. In this
system, the net magnetic charge is abelian as far as the first BPS
equation is concerned. The $\alpha$ monopole becomes nonabelian cloud
surrounding the massive $\beta$ monopoles. This is the case considered
in Ref.~\cite{dancer}. The position of the massless monopole form an
ellipsoid~\cite{lu1}. We expect that there are two independent
solutions for the second BPS equations in this case and that the
electric charges will depend on the the position of the apparently
massless $\alpha$ monopole if the gauge symmetry $SU(3)$ is broken to
$U(1)^2$. In string picture, one can see that the Bergman's
instability does not set in. Two D strings will connect three D3
branes in the letter V shape, emerging from the middle D3 branes and
the $\alpha$ monopole may appear massless because the $b_I$ vector
will emerge from the middle D3 brane to the middle point of the line
connecting two end D3 branes. F strings will connect the two tip D3
branes. Two tip D3 branes have the same coordinates in $b_I\phi_I$ and
so the $\alpha$ monopole appears massless. Note that the relation
between BPS dyons  in the field theory and D strings is
not direct in 1/4-BPS configurations.

In this letter, we have analyzed the role of massless monopoles in
1/4-BPS configurations, which are associated with multipronged
strings.  One insight obtained from this is that the positions of
massless monopoles can play the physical role.  There are a few
questions to remain.  We do not have a good understanding of some
cases. For example, consider a case where the gauge symmetry is
maximally broken to abelian subgroup, and some of magnetic monopoles
appear massless, but that the net magnetic charge is not purely
abelian. If such a case is possible, the question is that the global
color problem seems to appear. This may need more careful
understanding of the massless limit~\cite{lu2}. There has been several
works to understand the S-duality in case where there are nonabelian
massive magnetic monopoles~\cite{bais} and also the dual version of
massless monopoles. Our work may be helpful in this direction. For
this we need to understand how to quantize the classical
configurations we have obtained. Recently, the new way of obtaining
the electric charges has been obtained~\cite{tong}. It would be nice
to check our result by that method. Finally, we believe this work is a
step toward to the understanding of the attractive forces between
massive monopoles via massless monopoles.

\vspace{3mm} 

\centerline{\bf Acknowledgments} 

I thank Dongsu Bak for useful discussions. This paper is done
partially during the author's visit to Korean Institute for Advanced
Study. This work was also supported by KOSEP 1998 Interdisciplinary Research
Program and SRC program of SNU-CTP, and Ministry of Education BSRI
98-2418.

\end{document}